\documentclass{elsart}

\usepackage{natbib}

\usepackage{epsfig}

\usepackage{amssymb}

\begin{document}

\begin{frontmatter}



\title{GRB Observed by IBIS/PICsIT in the MeV Energy Range}


\author{V.~Bianchin}$^{a}$,
\ead{bianchin@iasfbo.inaf}
\author{L.~Foschini}$^{a}$,
\author{G.~Di~Cocco}$^{a}$,
\author{F.~Gianotti}$^{a}$,
\author{D.~G\"otz}$^{b}$,
\author{P.~Laurent}$^{b,c}$,
\author{G.~Malaguti}$^{a}$,
\author{F.~Schiavone}$^{a}$
\address{$^{a}$INAF/IASF-Bologna, via Gobetti 101, 40129 Bologna, Italy}
\address{$^{b}$CEA,IRFU, SAp, 91191 Gif sur Yvette, France}
\address{$^{c}$APC, B\^{a}timent Condorcet, 75205 Paris Cedex 13, France}

\begin{abstract}
We present the preliminary results of a systematic search for GRB and other 
transients in the publicly available data for the IBIS/PICsIT 
$(0.2-10~\rm{MeV})$ detector on board \emph{INTEGRAL}. 
Lightcurves in $2-8$ energy bands with time resolution from $1$ to 
$62.5~\rm{ms}$ have been collected and an analysis of spectral and temporal 
characteristics has been performed. This is the nucleus of a forthcoming 
first catalog of GRB observed by PICsIT. 

\end{abstract}


\end{frontmatter}

\parindent=0.5 cm

\section{Introduction}
PICsIT is the high-energy layer of the IBIS \citep{U03} instrument on-board 
the \emph{INTEGRAL} satellite, operating in the energy range between $200$~keV 
and $\sim 10$~MeV \citep{DC03}.
Scientific pointings are acquired in Standard Mode, which is the combination 
of the two sub-modes: Spectral Imaging and Spectral Timing (ST), optimized for 
spatial and temporal resolution, respectively. 
For both modes, events are integrated over time and energy intervals according 
to the on-board settings; energy bins are derived from the original $1024$ 
channels in conformity with the channel-to-energy relationship, defined in the 
on-board look-up tables (LUT) and derived from ground and in-flight 
calibration. 
ST mode allocates up to $8$ energy bins and time resolution down to 
$\sim 1$~ms. 
However the limited telemetry budget requires data to be integrated over the 
whole detector surface, so that the event spatial information is lost. 
The on-board configuration history for the ST mode can be found at the 
INAF/IASF-Bologna web page\footnote{http://www.iasfbo.inaf.it/extras/Research/INTEGRAL/Documentation/Hardware.html}.

\section{Method}
Spectral Timing data are processed by the specific software 
\texttt{OSA 7.0} \citep{G03}. 
Each lightcurve in output is then scanned to search for excesses in each 
energy band. 
Excesses are selected where the background subtracted count rate is above a 
significance threshold defined as $k*\sqrt{bkg}$, where $k=3$ above $360$~keV 
and it is set to $10$ below. 
The energy dependent threshold is applied to reduce spurious cosmic rays 
induced events which can mainly affect lightcurves below $\sim 300$~keV. 
The background is computed, for each energy band  and binning time, as the 
average rate over the pointing, neglecting saturation periods, telemetry 
gaps or oscillations at the Earth radiation belts passage. 
An excess is marked as a GRB candidate if the background corrected count rate 
remains above $1\sigma$ over a time interval of $5$~s around the peak. 
It is worth noting that with the present criteria only long GRB are selected 
and a different approach is to be implemented for the search of short events, 
which is hampered by the strong impact of cosmic rays background \citep{S03} 
and the lack of information on event position. 
Following the \emph{INTEGRAL} Burst Alert System (IBAS) approach \citep{M03}, 
the procedure applies to the original binning lightcurve and to rebinned 
lightcurves with time intervals of $0.5$, $1$ and $2$~s. 
The effect of increasing binning time is to dilute cosmic rays induced spikes 
which have a typical time scale of tens ms \citep{S03}. 
Since the procedure is still in a test phase in several cases the 
lightcurve is visually inspected. 

We cross-check PICsIT triggers with events of particle and solar origin, 
reported by the \emph{INTEGRAL} Radiation Environment Monitor 
(IREM - \citet{H03}) and by the Geostationary Operational Environmental 
Satellites (GOES)\footnote{http://www.swpc.noaa.gov}. 
The most stringent criterion to validate an excess as a GRB is based on the 
directional information of the event. 
Although PICsIT Spectral Imaging (SI) mode is optimized for spatial 
resolution, events of typical GRB time scale are not detected in SI data 
since these are on-board integrated over the pointing duration 
($\sim 2000$~s). 
In the present excess list only the very long ($\sim 160$~s) GRB~$041219$ was 
detected in both Spectral Imaging and Spectral Timing modes. 
Since the event position is not supplied by PICsIT Spectral Timing 
Mode data, to be confirmed as GRB, excesses are compared with GCN and other 
catalogs (KONUS-Wind, HETE, IBAS, IBIS/ISGRI). 

Finally we point out that the sky coverage of PICsIT increases with energy, 
since the passive collimator between the coded mask and IBIS shields the 
detector against the cosmic background up to $\sim 300$~keV and the active 
VETO surrounding the detector rejects spurious high-energy events up to 
$\sim 1$~MeV \citep{Q03}.
\section{Catalog status}
\begin{table}
\caption{GRB preliminary list. The table gives for each GRB: the start time 
of the event, the duration ($\Delta$T) in the PICsIT lightcurves, GCN number, 
the acquisition settings (time resolution and the number of energy bands), and 
the energy range (in keV) of the PICsIT detection.}
\vspace{0.2cm}

\begin{tabular}{ccccccc}
\hline
\hline
GRB & Time & $\Delta$T & GCN & Acquisition Settings       & Energy Range \\
    & UTC & (s)  &     & T bin~(ms) - En Bin Num & (keV)                 \\
\hline
\hline
030306 & 03:38:22 & 20 & 1930 & 62.5 - 2 & 208 - 676 \\
030307 & 14:32:00 &  5 & 1937 & 62.5 - 2 & 208 - 676 \\
030320 & 10:11:55 & 60 & 1941 & 62.5 - 2 & 208 - 676 \\
030405 & 02:17:28 & 10 & 2126 & 62.5 - 2 & 208 - 676 \\
030406 & 22:42:07 & 15 & 2127 & 62.5 - 2 & 208 - 676 \\
030422 & 07:51:15 & 15 & 2162 & 62.5 - 2 & 208 - 676 \\
041219 & 01:42:12 &160 & 2866 &    4 - 4 & 156 - 676 \\
050525A& 00:02:53 &  8 & 3466 &    4 - 4 & 260 - 676 \\
060901 & 18:43:51 & 10 & 5491 &    4 - 4 & 260 - 572 \\
061122 & 07:56:49 &  8 & 5834 &   16 - 8 & 208 - 572 \\
061222A& 03:28:52 & 15 & 5954 &   16 - 8 & 208 - 2600\\
\hline
\hline
\end{tabular}
\label{table1}
\end{table}
\begin{figure}
\begin{center}
\includegraphics[height=6.8cm,angle=-90]{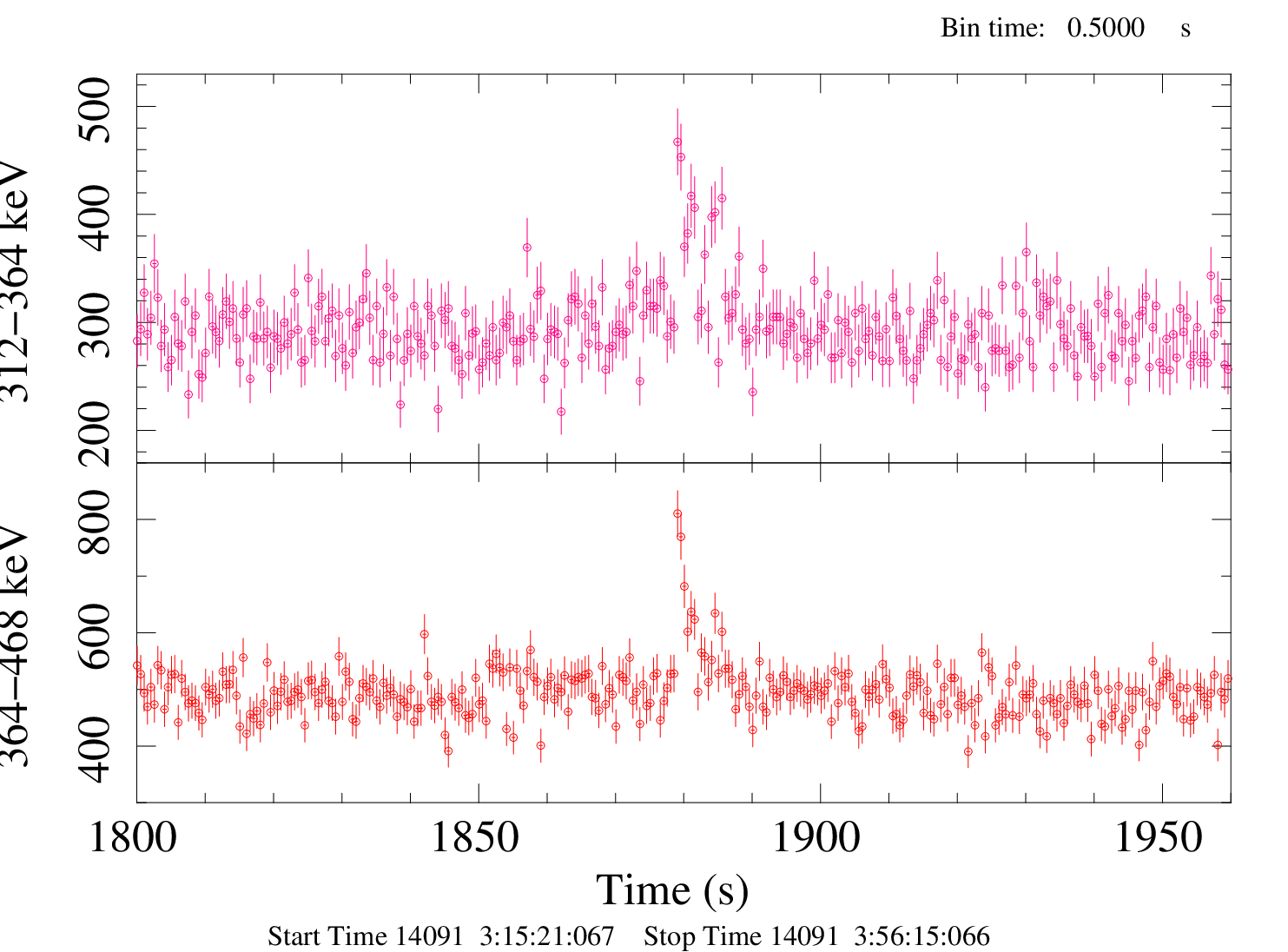}
\includegraphics[height=6.8cm,angle=-90]{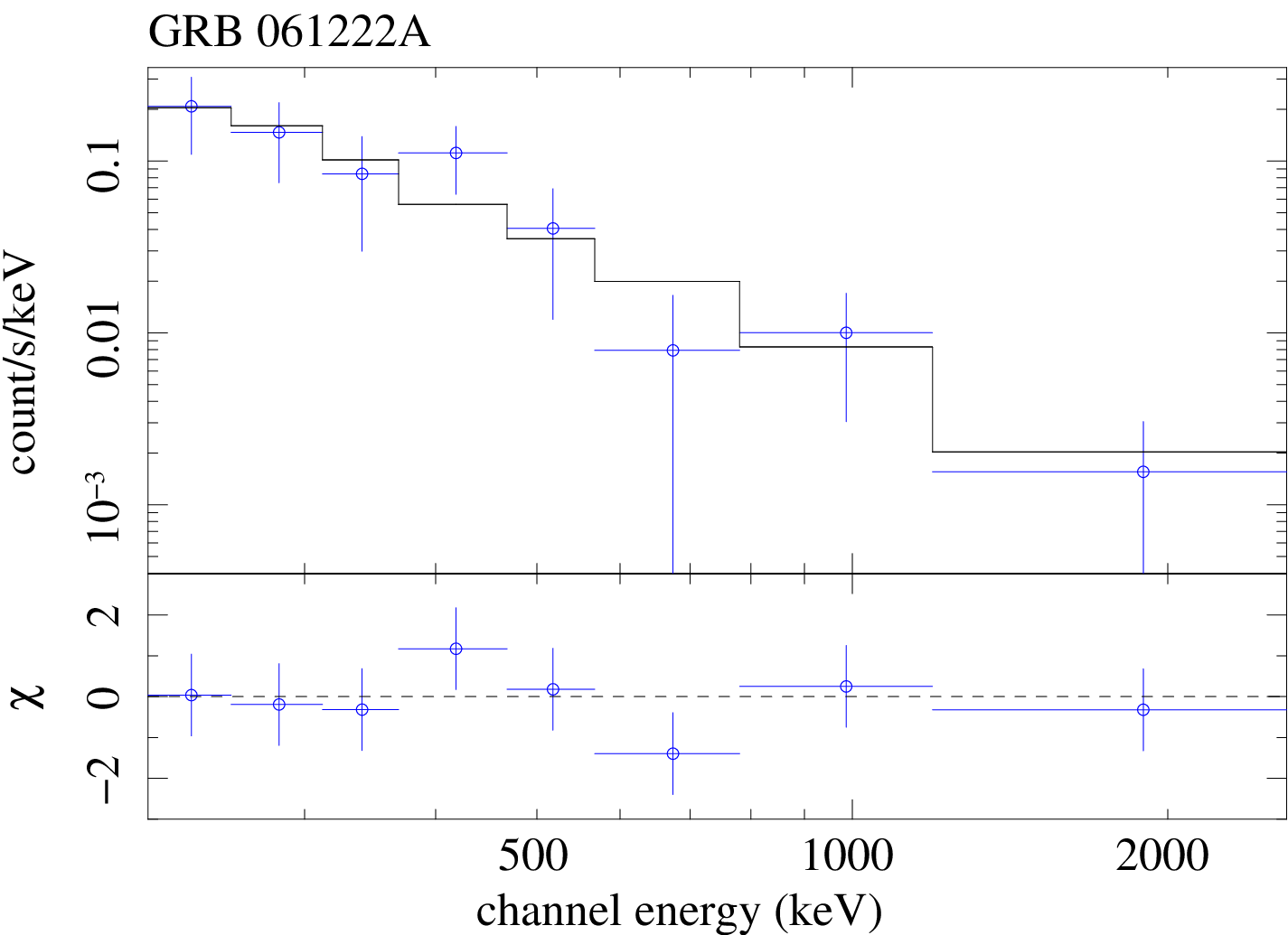}
\end{center}
\caption{\label{grb8}GRB~$061222$A lightcurve in the $312-364$ and 
$364-468$~keV energy bands (right panel) and spectrum in the energy range 
$208$~keV-$2.6$~MeV (left panel). This burst was not detected by IBIS/ISGRI.}
\end{figure}
\begin{figure}
\begin{center}
\includegraphics[width=5cm,angle=-90]{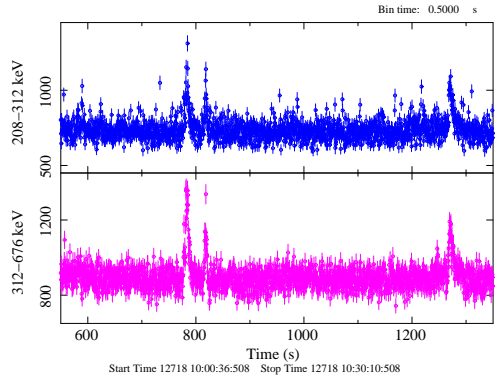}
\end{center}
\caption{\label{excess}The figure shows the double peaked GRB~$030320$ 
together with a remarkable excess occurred 8 minutes later, in the two energy 
bands $208-312$~keV and $312-676$~keV. The nature of this GRB-like excess is 
still not clear.}
\end{figure}
At present our sample covers $56$ complete revolutions (one \emph{INTEGRAL} 
orbit corresponds to $\sim 3$~days), from revolution $46$ to $68$, rev.~$311$ 
and from rev. $488$ to $520$. 
Moreover the sample includes single pointings corresponding to IBIS/ISGRI GRB 
detections. 
IBIS/PICsIT observed 11 GRB documented by GCN (Tab.~\ref{table1}) among which 
$7$ GRB are observed in $2$ energy bands up to $\sim 700$~keV, $2$ are 
observed in $4$ energy bands up to $\sim 700$~keV and one GRB (061222A) 
seems to extend up to the highest energy band ($1.2- 2.6$~MeV). 
In particular GRB~$061222$A satisfies the selection criteria in the energy range 
$312-572$~keV, however several counts, probably related to the event, are 
present in all energy bins. 

The updated list of events with lightcurves and spectra (if possible) is 
available at the web page\footnote{http://www.iasfbo.inaf.it/extras/Research/INTEGRAL/Catalogue/picsit\_soucat.html}. 
Spectral points are derived as the background subtracted count rate in each 
energy bin. 

Among the confirmed events, we mention the very long GRB~$041219$ \citep{M06} 
that was also detected in Spectral Imaging Mode and GRB~$061222$A \citep{B07} 
that appears up to $\sim 2.6$~MeV. 
In Fig.~\ref{grb8} the lightcurve in two energy bands ($312-364$ and 
$364-468$~keV) and the spectrum of GRB~$061222$A are shown. 
Despite the lightcurve suggests an energy dependent double peaked structure, 
the spectrum is integrated over the whole burst. 

Besides documented GRB we found several excesses from unknown sources, either 
Galactic Transients, or GRB-like peaks. 
The most intriguing case is the excess found on $2003$ March $20$ at 
$10:20:00$~(UTC), shown in Fig.\ref{excess} together with the documented 
GRB~$030320$ \citep{K03b}, occurring $\sim 500$~s  before, in the same 
pointing. 
The same excess is also present in the lightcurve\footnote{http://www.mpe.mpg.de/gamma/science/grb/1ACSburst/2003\_03\_main.html} of the Anti-Coincidence 
Shield SPI-ACS \citep{K03} however the imaging analysis with IBIS/ISGRI gives 
no detection. 
The satellite weekly report describes a nominal operational period, but no 
details about the IREM particle monitor are available. 
The solar activity was moderate-to-high over the whole week, however flares 
and particle events do not match with the excess. 
The event is not reported in any catalog and its nature is still 
under study. 

\section{Future work}
\begin{itemize}
\item The present sample of revolution will be extended to all public 
  pointings. 
\item The event catalog will include temporal and spectral information for 
  each excess. 
\item A procedure for the search of short and faint events will be 
  implemented. 
\end{itemize}

\end{document}